# How Does the Symmetry of $S_1$ Influence Exciton Transport in Conjugated Polymers?

[1]Raj Pandya*, [1]Antonios M. Alvertis, [1]Qifei Gu, [1]Jooyoung Sung, [2]Laurent Legrand, [3]David Kréher, [2]Thierry Barisien, [2]Alex W. Chin, [1]Christoph Schnedermann, and [1]Akshay Rao*

[1]Cavendish Laboratory, University of Cambridge, J.J. Thomson Avenue, CB3 0HE, Cambridge, United Kingdom

[2]Sorbonne Université, CNRS, Institut des NanoSciences de Paris, INSP, 4 place Jussieu, F-75005 Paris, France

[3]Sorbonne Université, CNRS, Institut Parisien de Chimie Moléculaire (IPCM) UMR 8232, Chimie des Polymères, 4 Place Jussieu, 75005 Paris, France

correspondance : *rp558@cam.ac.uk, ar525@cam.ac.uk

**Abstract**

Many optoelectronic devices based on organic materials require rapid and long-range singlet exciton transport. Key factors that control the transport of singlet excitons includes the electronic structure of the material, disorder and exciton-phonon coupling. An important parameter whose influence on exciton transport has not been explored is the symmetry of the singlet electronic state ($S_1$). Here, we employ femtosecond transient absorption spectroscopy and microscopy to reveal the relationship between the symmetry of $S_1$ and exciton transport in highly aligned, near-disorder free, one-dimensional conjugated polymers based on polydiacetylene. By altering the torsional angle of the polymer backbone we control the symmetry of the $S_1$ exciton and find that excitons with $2^1A_g^-$ symmetry exhibit a significantly higher diffusion coefficient ($34 \pm 10$ cm$^2$ s$^{-1}$) compared to excitons with $1^1B_u^+$ symmetry ($7 \pm 6$ cm$^2$ s$^{-1}$). We also find that while exciton transport in the $2^1A_g^-$ state occurs without exciton-exciton annihilation effects, $1^1B_u^+$ excitons undergo exciton-exciton annihilation. Both states are found to exhibit sub-diffusive behaviour, despite the large diffusion coefficient. *Ab initio GW*-BSE calculations reveal that this is due to the comparable strengths of the exciton-phonon interaction – which tends to localise excitons, and the exciton coupling – which tends to delocalise excitons. Our results demonstrate an intricate link between electronic state symmetry and exciton transport in π-conjugated polymer systems and also highlight the role of exciton-phonon coupling.

Organic optoelectronic devices ranging from light-emitting diodes to photovoltaic cells and transistors[1–3] are frequently based on linear polymers with a $C_{2h}$ point group symmetry. Here, the electronic ground state ($S_0$) exhibits $1^1A_g^+$ symmetry. Depending on the polymer conjugation length and backbone geometry the first ($S_1$) and second ($S_2$) excited electronic states are either of $2^1A_g^-$ or $1^1B_u^+$ symmetry[4,5]. Irrespective of the exact state ordering, photoexcitations in conjugated polymers rapidly forms excitons in the lowest energy excited $S_1$ state. When the $S_1$ state is of $2^1A_g^-$ symmetry the materials are typically non-fluorescent, the $S_1$ state supports some triplet-pair character $^1(TT)$[6–8] and frequently shows a short electronic lifetime. On-the-other hand polymers with an $S_1$ state of $1^1B_u^+$ character are often luminescent with long electronic lifetimes[9,10].

Despite our considerable understanding of the electronic structure of conjugated polymers[3,4,11–15], the impact of the electronic symmetry of the $S_1$ state on exciton diffusion behaviour remains largely unexplored. In other words, is it not known whether a $2^1A_g^-$ or $1^1B_u^+$ symmetry state would provide better exciton transport properties? At first glance it might appear that the longer lifetimes of $1^1B_u^+$ states would enable longer exciton diffusion lengths. But the significantly altered electronic structure and many-body character of $2^1A_g^-$ states may provide access to unique diffusion properties. The experimental study of exciton diffusion in molecular systems is challenging due to the difficulties associated with measuring ultrafast, nanoscale exciton diffusion properties in systems that exhibit sub-100 ps electronic lifetimes coupled with sub-100 nm exciton diffusion lengths[12,13]. Critically challenges associated with systematically tuning polymer electronic structure without introducing detrimental effects such as intrachain disorder, cross-linking, *etc.* have until now prevented a systematic investigation of electronic symmetry on exciton diffusion.

Here, we overcome these problems by using femtosecond transient absorption spectroscopy and microscopy[20–23] to measure exciton transport in different topochemically (*i.e.* near defect-free as compared to solution processed materials) polymerised polydiacetylene (PDA) chains. PDAs offer an ideal platform to study the effect of electronic properties on exciton transport for several key reasons. Structurally, PDAs consist of extended (2 to 5 µm) one-dimensional polymer chains that are distinguished by the degree of back-bone torsion. Perfectly planar polymer chains are typically referred to as 'blue' PDA, while polymer chains with a defined back-bone torsional angle (14°) are known as 'red' PDA. Other than the backbone torsion both PDAs show near-identical structural and chemical material properties, *i.e.* degrees of polymer content, polymer alignment, structural disorder, *etc.* (SI, Figure S1). Electronically, both PDA phases display strong coupling between the monomer units and exhibit large exciton coherence lengths[18,24–26]. Owing to the planar backbone geometry, 'blue' PDA exhibits an optically accessible $1^1B_u^+$ ($S_2$) which lies above a $2^1A_g^-$ ($S_1$) state. This ordering gives rise to a broad visible absorption spectrum dominated by vibrational progressions at 1400-1500 and 2100 cm$^{-1}$ (C=C and C≡C stretching mode, blue spectrum; Figure 1b). By contrast, the energetic ordering in 'red' PDA is reversed due to backbone torsions, resulting in a dark $2^1A_g^-$ ($S_2$) state lying above the optically-accessible $1^1B_u^+$ ($S_1$) state. As a consequence, 'red' PDA exhibits a significantly blue-shifted absorption spectrum compared to 'blue' PDA (Figure 1b).

The ensemble excited state properties of 'blue' and 'red' PDA films following photo-excitation with a 10 fs pump pulse centred at 560 nm[27] are shown Figure 1c,d. In 'blue' PDA, the pump-probe spectrum is characterised by a sharp stimulated emission (SE) band at 650 nm, a broad photoinduced absorption (PIA) at 750 – 900 nm and a second narrower PIA at 680 – 700 nm (Figure 1c). By fitting the electronic decays (exponential decay convoluted with instrument response) we find that the SE decays mono-exponentially with a lifetime of 630 ± 10 fs, while the PIA at 750 – 900 nm decays with a lifetime 1300 ± 40 fs and the PIA at 680 – 700 nm with 2300 ± 130 fs. The observed excited-state features and dynamics agree excellently with previous reports[26,28,29]. Following this agreement, we can assign the SE band to $1^1B_u^+$ ($S_2$), the broad PIA is attributed to the $2^1A_g^-$ ($S_1$) state and the narrow PIA at 680 – 700 nm to a hot, vibrationally excited, ground state[29–31]. Photoexcitation of 'blue' PDA to $1^1B_u^+$ ($S_2$) is thus followed by rapid internal conversion within 90 fs to $2^1A_g^-$ ($S_1$), as determined by spectral decomposition methods[28], which decays back to the ground state. In 'red' PDA the transient absorption spectrum shows a single broad PIA in the 650 – 900 nm range. The decay of this is PIA is 13.8 ± 0.2 ps and has been suggested to correspond to the transition between $1^1B_u^+$ (which is $S_1$ state in this system) and a state lying at ~3-3.5 eV, of $^nA_g$ or charge transfer (CT) character[24]. We emphasise that we carried out both fs-TA (and fs-TAM measurements, see below) as a function of carrier density (initial carrier concentration ($n_0$) in a range from $n_0 = 7.1 \times 10^{16} – 1.3 \times 10^{18}$ cm$^{-3}$; SI, S4). In both cases these results showed that for 'blue' PDA the kinetics do not vary with carrier concentration whereas annihilation was observed for 'red' PDA.

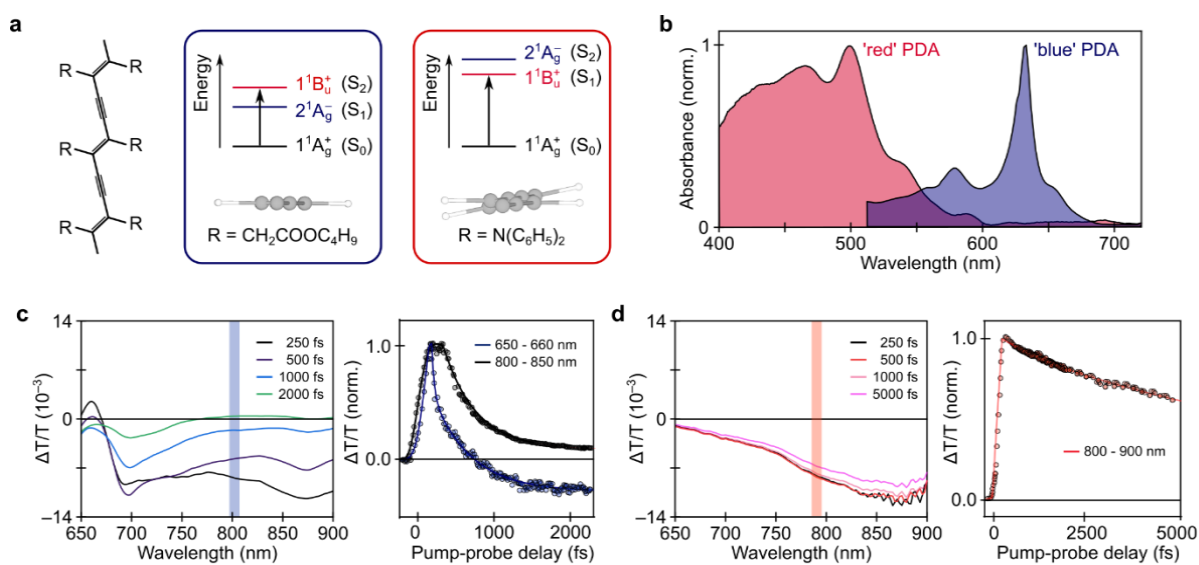

**Figure 1: Structural and optical characterisation of 'blue' and 'red' PDA. a.** Chemical structure and packing side-chains of 'blue' (left) and 'red' (right) PDA. In 'blue' PDA the backbone is planar,

whereas in 'red' PDA there is 14° twist angle between adjacent monomer units in the chain. **b.** Absorption spectra of 'blue' and 'red' PDA. **c-d.** Selected spectral slices of pump-probe spectra of 'blue' (c) and 'red' PDA (d). The photoinduced absorption bands of the $2^1A_g^-$ and $1^1B_u^+$ states are marked. Shaded regions indicate probe wavelengths used in transient absorption microscopy experiments. The right panels show kinetics and associated exponential decay fits at indicated probe wavelengths.

In order to extract the exciton diffusion characteristics of 'blue' and 'red' PDA we performed widefield femtosecond transient absorption microscopy (fs-TAM) with a similar 10 fs pump pulse centred at 560 nm[21,23,32]. In fs-TAM a pump pulse is focussed *via* a high-numerical aperture (NA=1.1) objective to near the diffraction limit Gaussian spot ($\sigma$ = 143 nm), which locally creates excitons within the sample. After a variable time delay, a broadband, 7 fs probe pulse centered at 780 nm is loosely focussed on the sample to image the pump-induced signal changes in the material. The transmitted probe is subsequently imaged onto a two-dimensional detector. By recording the transmission image with and without the pump pulse incident on the sample, we calculated the spatially-resolved transient absorption image of the locally excited exciton distribution as a function of pump-probe time delay. Tracking its temporal evolution then allows to extract information about exciton diffusion properties. For 'blue' PDA we probed in the centre of the $2^1A_g^-$ PIA band at 800 nm (blue stripes in Figure 1c), while the probe wavelength in 'red' PDA was set to 790 nm to be centred with the $1^1B_u^+$ PIA (orange stripes in Figure 1d). These wavelengths ensure that we exclusively monitor the $S_1$ exciton population, free of other spectral features.

To quantify the spatial extent of the signal, we extracted the spatial standard deviation for all transient absorption images using a 2D Gaussian fit (see SI, S3). We subsequently computed the mean-square-displacement (MSD) for the underlying spatial exciton profile, which is a measure of the change in the spatial extent of the exciton population as a function of time. As shown in Figure 2a,b, both samples exhibit a pronounced increase in their MSD over the first 5 ps with 'blue' PDA displaying MSD values exceeding 'red' PDA. In both PDA phases, the exciton propagation is highly anisotropic and proceeds exclusively along a single axis (see SI, S3). This behaviour is expected based on the one-dimensional structure of PDA and the low polymer content ($10^{-4} – 10^{-3}$ w/w) of the samples remains, preventing intrachain transport. Our results therefore suggest that exciton propagation proceeds along the long-axis of the PDA chains (monoclinic *b* axis of crystals)[24,26]. We remark that varying the probe wavelength within the broad PIA did not alter the observed transport behaviour (SI, Figure S2).

Although the PDA chains are highly ordered, their thin films are known to contain some underlying nanoscale inhomogeneity. We therefore repeated fs-TAM measurements on 30 sample locations in 'blue' PDA and 21 in 'red' PDA, to build up a statistical picture of singlet exciton transport.

Furthermore, to accurately compare the spatial dynamics of both samples, we restricted our analysis to time-delays >250 fs in order to avoid effects due to $S_2$-$S_1$ internal conversion processes in 'blue' PDA.

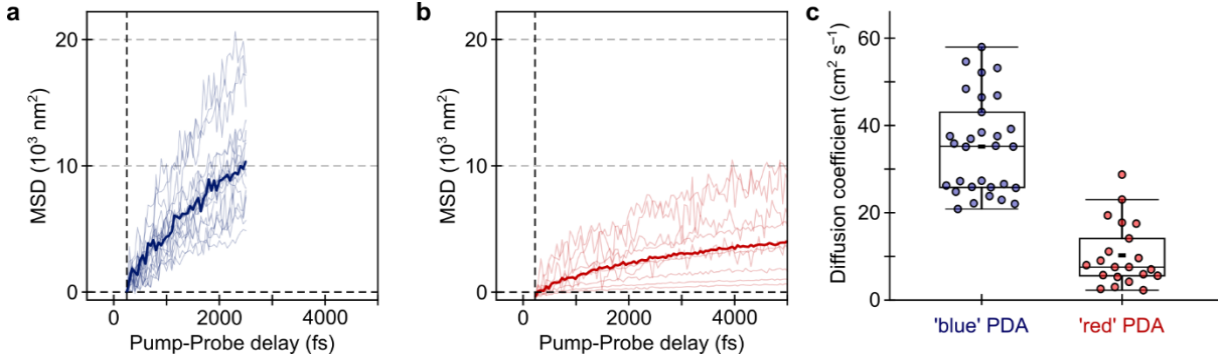

**Figure 2: Femtosecond transient absorption microscopy results for 'blue' and 'red' PDA. a-b.** MSD curves for 'blue' (a) and 'red' PDA (b), respectively. Solid line shows average curves whereas faint lines are from individual sample locations. Due to the faster decay in 'blue' PDA, the MSD was only recorded to 2.3 ps. **c.** Scatter-box plot of diffusion coefficients for 'blue' and 'red' PDA. Boxed area is 25-75% range, with horizontal line representing median and whiskers for interquartile range. Filled black square represent the mean.

A salient feature of the MSD curves displayed in Figure 2a and 2b is that for 'blue' PDA the maximum MSD value reached is typically greater than 'red' PDA. In order to extract the relevant diffusion coefficients, we constructed the associated differential diffusion equation[33]

$$\frac{\partial n(x,t)}{\partial t} = D \left[\frac{\partial^2 n(x,t)}{\partial x^2}\right] - \frac{n(x,t)}{\tau} - \frac{\gamma\, n(x,t)^2}{\sqrt{\tau}}. \qquad (1)$$

Here $n(x,t)$ is the electronic population at a time $t$ and position $x$, $D$ is the diffusion coefficient, $\tau$ is the electronic lifetime and $\gamma$ is a coefficient which accounts for exciton-exciton annihilation. In the absence of exciton annihilation this equation can be solved to show that in one dimension,

$$\mathrm{MSD}(t) = A\,(t - t_0)^\alpha, \qquad (2)$$

where $A$ is proportionality factor and α accounts for non-linearity in time. In the case of α = 1 we retrieve normal diffusion, while α > 1 is known as super-diffusive behavior and α < 1 is referred to as sub-diffusive behavior. In this latter case a time-dependent diffusivity $D(t)$ can be defined as,

$$D(t) = \tfrac{1}{2} A\, t^{\alpha - 1}. \qquad (3)$$

In the presence of annihilation no analytical solution exists and equation (1) must be fit to extract $D$ and $\gamma$ [33] (see SI, S3 for further details).

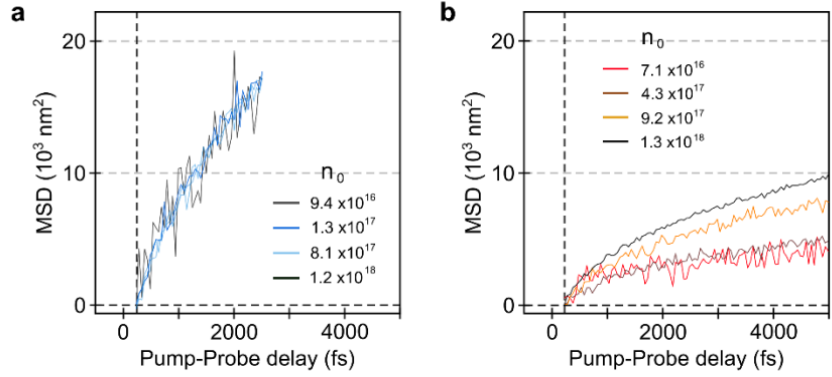

**Figure 3: Femtosecond transient absorption microscopy as a function of carrier density for 'blue' and 'red' PDA. a-b.** In the case of 'blue' PDA (a) there is no dependence of the MSD on the carrier density, whereas in 'red' PDA (b) the diffusion coefficient increases at higher pump fluences. In this latter case exciton-exciton annihilation effects can be considered to be playing a role.

To identify, which model should be applied, we carried out a fluence dependence on the same sample spot for both PDAs, as shown in Figure 3. Here, we find that the transient spatio-temporal dynamics of 'blue' PDA do not vary with power (Figure 3a), while 'red' PDA displays a pronounced power dependence (Figure 3b), indicative of a higher annihilation rate. This increased annihilation rate of $1^1B_u^+$ compared to $2^1A_g^-$ excitons is an important observation that is not well understood. While beyond the scope of this article, we note that the $2^1A_g^-$ state exhibits a lower transition dipole moment than $1^1B_u^+$ [7,34]. Given that the annihilation cross-section of a state is proportional to its transition dipole moment[35–37] this may explain why the $2^1A_g^-$ symmetry excitons do not exhibit significant annihilation effects[38]. The $2^1A_g^-$ excitons can still scatter in such a way that would lead to annihilation (impact ionisation, Auger effects, *etc*), but this would require the excitons to be very close, or at high densities. We note that it is possible that local geometry/structure around the excitons prevents them from approaching each other, and that the excitation density does not reach the magnitudes required for short-range annihilation mechanisms to be operative (on the timescales of internal conversion).

Based on the observed power dependence, we applied equation 1 to directly extract diffusion consistent, $D$, and annihilation parameter, $\gamma$ for 'red' PDA. In contrast, for 'blue' PDA we fit equation 2 to the MSD trace to extract $A$ and $\alpha$ and then extract the diffusion coefficient from the time-dependent diffusivity $D(t)$ according to equation 3. Following this analysis, we derive a mean diffusion coefficient of $34 \pm 10$ cm$^2$ s$^{-1}$ for 'blue' PDA (value at 2500 fs, see SI, S3 for full $D(t)$), which is significantly larger and more varying across different sample locations as compared to 'red' PDA with $7 \pm 6$ cm$^2$ s$^{-1}$. We remark that the spread in diffusion coefficients for 'blue' PDA is larger than for 'red'. This tentatively suggests that diffusion in 'red' PDA is less susceptible to effects of sample morphology but requires further studies. The exponent value, α, for 'blue' PDA lies between 0.7 – 0.9 and indicates sub-diffusive exciton transport. For 'red' PDA the annihilation coefficient, $\gamma$, ranged between 0.3 – 1 cm$^2$ s$^{-1}$ (SI, S5). Critically, while this annihilation factor is substantial, the ratio of the mean diffusion coefficient to mean annihilation coefficient is 17.5. This suggests that the annihilation is still only a minor contributor to the sub-diffusive transport[39].

Having discussed the origin of the observed exciton dynamics we can determine the overall diffusion length from

$$L_D = \sqrt{2D\tau_{elec}}, \qquad (5)$$

where $\tau_{elec}$ is the intrinsic lifetime of the underlying process. Previously we have determined for $2^1A_g^-$, $\tau_{elec}$ is ~600 fs, whereas for $1^1B_u^+$, $\tau$ ~9000 fs [28]. Based on these values we can extract diffusion lengths of 26 ± 6 nm and 33 ± 5 nm, for 'blue' and 'red' PDA, respectively. It is noteworthy to mention that the diffusion length is similar for both systems, despite their stark difference in diffusion coefficient and annihilation characteristics. Long-range exciton transport thus requires not only highly mobile excitons with minimized annihilation, but also long state lifetimes.

Despite the high diffusion coefficients obtained at room temperature in these systems (Figure 2), which would be suggestive of a strong tendency of excitons to delocalize, we observe sub-diffusive exciton transport in both $2^1A_g^-$ and $1^1B_u^+$ states, which is typical of strong exciton-phonon interactions[40] that lead to localized excitons. In order to understand these seemingly inconsistent observations, we employ *ab initio GW*-BSE calculations to quantify the two competing effects. The electronic coupling between excitons residing on neighbouring monomers of the polymer chain (*J*-coupling) drives the system to a more delocalized state. From a calculation of the exciton bandwidth (the range of the energy-momentum dispersion along the chain, SI, S6) we could determine $J$ = 0.33 eV for 'red' PDA ($1^1B_u^+$ state). Unfortunately, the multi-excitonic character in 'blue' PDA ($2^1A_g^-$ state) prevents us from obtaining the corresponding *J* value within theories that capture single excitations. However, we can estimate the *J*-coupling of 'blue' PDA to be larger than 0.33 eV (SI, S6 for further details), since the more planar backbone of 'blue' PDA as compared to 'red' PDA increases the *p*-orbital overlap, and thus the π-conjugation along the polymer chains in the system. Moreover, we quantify the strength of exciton-phonon coupling by calculating the reorganization energy of the $1^1B_u^+$ state of PDA to be λ = 0.43 eV (SI, S7), which denotes the driving force of the system to localize after photoexcitation.

Based on the reorganization energy and *J*-coupling strength, we can now classify the transport regime in which PDAs operate. If *J*-coupling is small compared to the reorganization energy, we anticipate the system to localize on a single monomer and transport will only occur via incoherent hopping. Conversely, in our PDA films, we calculated, $J \sim \lambda$. This suggests that exciton transport may operate in the coherent regime where exciton delocalization is dominant[41,42]. However, the similarity between *J* and λ also suggests that exciton localization via exciton-phonon couplings can still occur. The subtle interplay between exciton-phonon coupling (localization) and *J*-coupling (delocalization) demonstrates that exciton transport occurs through the motion of partially delocalized excitons that reside on several monomers, as visualised in Figure 4. This leads to sub-diffusive exciton transport with high diffusion coefficients.

The larger *J*-coupling strength of 'blue' PDA ($2^1A_g^-$) compared to 'red' PDA ($1^1B_u^+$) suggests an enhanced delocalization and thus an increased diffusion coefficient, as observed experimentally (Figure 2). We also highlight that an additional factor that may play a role in the greater diffusion coefficient of 'blue' PDA ($2^1A_g^-$) is that in contrast to 'red' PDA ($1^1B_u^+$), $2^1A_g^-$ is a superposition state with many-body (CT, triplet-pair, *etc*) character[7,28]. Following photoexcitation, the initial $2^1A_g^-$ state may rapidly decay into one of these states with a greater spatial extent *e.g.* triplet-pairs. This state may be more spatially delocalized and less susceptible to annihilation. We note that in the more delocalized, spatially separated $2^1A_g^-$ state, dipolar interaction between excitons, required for annihilation, will be weaker further reducing this effect.

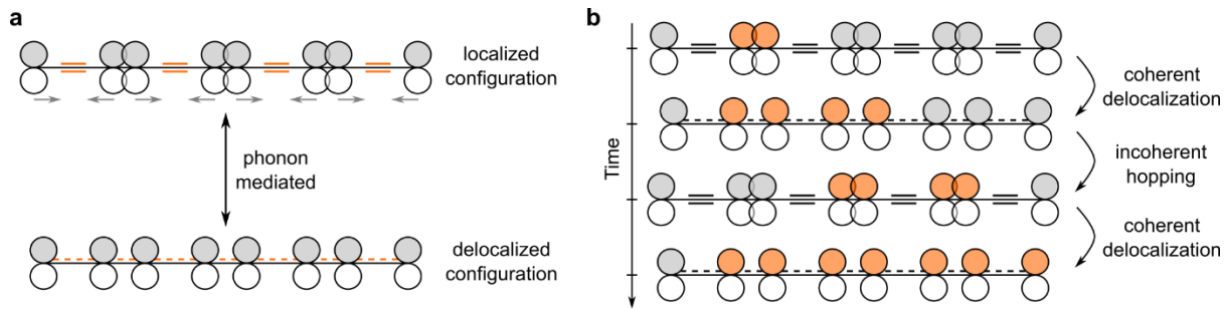

**Figure 4: Cartoon schematic demonstrating the interplay of localizing and delocalizing interactions**. **a.** The large *J*-couplings result in delocalization of the excitons, but strong-exciton phonon coupling causes localization. **b.** This interplay between *J*-coupling and exciton-phonon interactions results in sub-diffusive transport (hopping), but with a high diffusion coefficient. The motion of monomer units is only for illustrative purposes.

Furthermore, an additional factor that may play a role in the greater diffusion coefficient of 'blue' PDA ($2^1A_g^-$) is that in contrast to 'red' PDA ($1^1B_u^+$), $2^1A_g^-$ is a superposition state with many-body (CT, triplet-pair, *etc*) character[7,28]. Following photoexcitation, the initial $2^1A_g^-$ state may rapidly decay into one of these states with a greater spatial extent *e.g.* triplet-pairs. This state may be more spatially delocalized and less susceptible to annihilation. We note that in the more delocalized, spatially separated $2^1A_g^-$ state, dipolar interaction between excitons, required for annihilation, will be weaker further reducing this effect.

**Conclusion**

In summary we have investigated the influence of electronic state symmetry on exciton diffusion in conjugated polymers. We have shown $2^1A_g^-$ excitons have on average ~3 times higher diffusion coefficients than $1^1B_u^+$ excitons in PDA polymer chains. Where the exciton transport of $1^1B_u^+$ excitons exhibits annihilation effects, for $2^1A_g^-$ excitons, the small transition dipole moment and many-body character results in no measurable annihilation. The motion of both $2^1A_g^-$ and $1^1B_u^+$ excitons however appears to remain sub-diffusive likely due to exciton-phonon mediated trapping effects. Despite its smaller diffusion coefficient, the longer electronic lifetime of $1^1B_u^+$ means that $2^1A_g^-$ and $1^1B_u^+$ excitons have a similar diffusion length (~30 nm). Our results suggest that $2^1A_g^-$ excitons are able to move across space as effectively as excitations that live orders of magnitude longer. This is in part due to the absence of annihilation losses.

Sample preparation

'Blue' Polydiacetylene

3BCMU (3-methyl-n-butoxy-carbonylmethyl-urethane) diacetylene molecules were synthetized in-house using the method previously outlined by Se *et al.* and references therein[43]. The synthesis classically consisted of two steps: (i) oxidative coupling of 4-pentyn-1-ol (Hay's method) to produce the 4,6-decadiyn-1,10-diol and then, (ii) reaction of the diol with *n*-butylisocyanate acetate. Note that the source 4-pentyn-1-ol was not synthetized at the laboratory but purchased from Sigma-Aldrich (Merck). Ultrathin single crystals were grown between two coverslips using a melt-processing method. The whole process was systematically carried out under a polarized optical microscope so as to be able to follow and control the sample elaboration: a very small amount of diacetylene powder is placed at one edge of the double-slides assembly; when heating above the melting temperature (~65°C) the liquid diacetylene fills the empty space by capillary action to form a thin liquid film between the two substrates. Rapid cooling leads to the formation of a highly polycrystalline film. The sample is then heated again to around the melting temperature until the melting of all the crystallites took place. When only a few crystal germs remain the sample is cooled again at a very slow cooling rate (typically < 0.1°C/mn) to induce the growth of large single monocrystalline domains from the germs. Typical polymer contents by weight are then in the $10^{-4}$ - $10^{-3}$ range. The intrachain separation (~100 nm; homogeneous distribution) is obtained from the absorption optical density ($OD$) and Beer Law, where $OD = \frac{\propto \times l \times c}{2.3}$ where $\propto$ is taken as $\sim 1 \times 10^6$ cm$^{-1}$ for light polarized parallel to the long axis of the chains and $l$ is ~150 μm. The spatial separation between chains is assumed to be significantly large enough that there are negligible inter-chain interactions.

'Red' Polydiacetylene

To make multi-pulses microscopy possible by limiting scattering of the pump pulse ultra-thin crystalline monodomains of the ('red' PDA: 3Nϕ$_2$) diacetylene were grown before being slightly polymerized. Large size crystalline regions (~ mm$^2$) with sub-micronic thicknesses (~ 100 - 200 nm) are obtained using a melt processing method[44]. A few mg of the purified diacetylene powder are melt (T ≥ 80 °C) and the liquid is injected, using capillarity action, in the space between two superimposed microscope coverslips. After rapid cooling the thin liquid film crystallization leads to the formation of a highly polycrystalline structure, being the assembly of microdomains. The sample is then heated again under a polarized optical microscope until almost all the melting takes place but preserving a few single crystal germs. At that point the sample is cooled again at a very slow cooling rate to induce the growth of hundreds of μm – mm scale domains from the germs until room temperature is reached. The (3Nϕ$_2$) is a highly stable monomer that is characterized by both weak thermal- and weak UV photo- reactivity[24]. Polymer chains are thus generated by exposure to X-ray light in a diffractometer (Rigaku, Smartlab). The polymer content is adjusted by trial and error and kept low enough (typically below 0.1 % in weight) to form a solid solution of isolated polydiacetylene chains inside their diacetylene host crystal.

Absorption spectroscopy

Polarised absorption spectroscopy of PDA crystals was performed using a home built setup with a white light source generated by focussing the fundamental of a Yb-based amplified system (PHAROS, Light Conversion) into a 4 mm YAG crystal. The resulting absorption (corrected for the sample substrate) was then collected by imaging with a Silicon photodiode array camera (Entwicklunsbüro Stresing; visible monochromator 550 nm blazed grating).

Femtosecond pump-probe spectroscopy

The fs-TA experiments were performed using a Yb-based amplified system (PHAROS, Light Conversion) providing 14.5 W at 1030 nm and 38 kHz repetition rate. The probe beam was generated by focusing a portion of the fundamental in a 4 mm YAG substrate and spanned from 520 nm to 1400 nm. The pump pulses were generated in home-built noncollinear optical parametric amplifiers (NOPAs), as previously outlined by Liebel *et al.*. The NOPAs output (~4 to 5 mW) was centred typically between 520 and 560 nm (FWHM ~65-80 nm), and pulses were compressed using a chirped mirror and wedge prism (Layerterc) combination to a temporal duration of ~9 fs. Compression was determined by second-harmonic generation frequency-resolved optical gating (SHG-FROG; upper limit) and further confirmed by reference measurements on acetonitrile where the 2200 cm$^{-1}$ mode could be resolved. The probe white light was delayed using a computer-controlled piezoelectric translation stage (Physik Instrumente), and a sequence of probe pulses with and without pump was generated using a chopper wheel (Thorlabs) on the pump beam. The pump irradiance was set to a maximum of 38 μJ/cm$^2$. After the sample, the probe pulse was split with a 950 nm dichroic mirror (Thorlabs). The visible part (520–950 nm) was then imaged with a Silicon photodiode array camera (Entwicklunsbüro Stresing; visible monochromator 550 nm blazed grating). The near infrared part was imaged using an InGaAs photodiode array camera (Sensors Unlimited; 1200 nm blazed grating). Measurements were carried out with a time step size of 4 fs out to 2 ps to minimize the exposure time of the sample to the beam. Unless otherwise stated, all measurements were carried out with the probe polarisation set parallel with respect to that of the pump (using a half-waveplate; Eksma) and along the PDA chains.

The absorption spectrum of samples was measured after each pump-probe sweep to account for any sample degradation.

Femtosecond pump-probe microscopy

The femtosecond wide-field detected transient absorption microscope has been described in detail previously. Briefly, a Yb:KGW amplifier system (LightConversion, Pharos, 5 W, 180 fs, 1030 nm, 200 kHz) was used to seed two white-light stages for pump and probe generation. The pump white-light (3 mm Sapphire) was spectrally adjusted with a 650 nm short-pass filter (FESH650, Thorlabs), and compressed to 10 fs for all optical elements with two pairs of third-order compensated chirped mirrors and a wedge-prism pair (Layertec). Subsequently, the mode of the pump pulse is cleaned by a pinhole before being focused through the objective lens (NA = 1.1, oil immersion) to a spot size of ~260 nm (full-width-half-maximum). The probe white-light (3 mm YAG) was spectrally adjusted to 650 – 900 nm in a home-build fused silica prism filter and compressed to 7 fs with a pair of third-order compensated chirped mirrors and a wedge-prism pair (Venteon) before being free-space focused onto the sample (20 micron Gaussian spot size full-width-half-maximum). The transmitted probe was imaged onto an emCCD (Rolera Thunder, Photometrics) at 55.5 nm/pixel as verified by a resolution target. The frame rate of the camera was set to 30 Hz with an integration time of 11 ms and pump off/on images were generated by a mechanical chopper at a frequency of 15 Hz. For the measurements, we adjusted the pump fluence to achieve initial carrier concentrations ($n_0$) in the range: $n_0 = 7.1 \times 10^{16} - 1.3 \times 10^{18}$ cm$^{-3}$ for all samples.

Supporting Information

Acknowledgements

We acknowledge financial support from the EPSRC and Winton Program for the Physics of Sustainability. R.P. additionally thanks the EPSRC for a Doctoral Prize Fellowship, Andrew Musser (Cornell University, USA) for useful initial discussions, and Lily Russell-Jones (London School of Economics) and Matthew Storer (London) for invaluable guidance. A.M.A. acknowledges the support of the Engineering and Physical Sciences Research Council (EPSRC) for funding under grant EP/L015552/1. C.S. acknowledges financial support by the Royal Commission of the Exhibition of 1851.

Associated Content

The data associated with this manuscript is freely available at [url to be added in proof].

For table of contents only

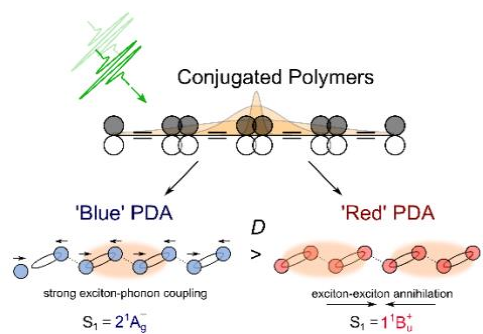